**Title**

Probabilities of causation and post-infection outcomes

**Authors**

Bronner P. Gonçalves[1]

**Affiliations**

[1] Faculty of Health and Medical Sciences, University of Surrey, 30 Priestley Rd, Guildford GU2 7YH, United Kingdom

**Correspondence**

Faculty of Health and Medical Sciences, University of Surrey, 30 Priestley Rd, Guildford GU2 7YH, United Kingdom, b.goncalves@surrey.ac.uk and bronnergoncalves@gmail.com

**Abstract**

Probabilities of causation provide explanatory information on the observed occurrence (causal necessity) and non-occurrence (causal sufficiency) of events. Here, we adapt these probabilities (probability of necessity, probability of sufficiency, and probability of necessity and sufficiency) to an important class of epidemiologic outcomes, post-infection outcomes. A defining feature of studies on these outcomes is that they account for the post-treatment variable, infection acquisition, which means that, for individuals who remain uninfected, the outcome is not defined. Following previous work by Hudgens and Halloran, we describe analyses of post-infection outcomes using the principal stratification framework, and then derive expressions for the probabilities of causation in terms of principal strata-related parameters. Finally, we show that these expressions provide insights into the contributions of different processes (absence or occurrence of infection, and disease severity), implicitly encoded in the definition of the outcome, to causation.

*Introduction*

Causal research in epidemiology typically focuses on "summary" (e.g. average) effects [1, 2]. In some circumstances, however, it might be useful to study causal attribution [3]. The probability of necessity ($p_N$) [4], an estimand that requires evaluating a counterfactual outcome under absence of exposure conditional on the observed presence of both outcome and exposure, is directly relevant to this type of investigation. Other probabilities (probability of sufficiency and probability of necessity and sufficiency) have also been described that capture other aspects of causation [5] and that are related to $p_N$.

Here, our objective was to adapt these probabilities of causation to research on post-infection outcomes. Briefly (a more detailed discussion is provided below), these are studies that, in defining the outcome of interest, explicitly account for infection acquision, and for participants who do not become infected, the outcome is undefined; in other words, analogous to settings with truncation by death [6], these analyses condition on a post-treatment variable. In addition to defining the probabilities of causation for this class of outcomes, we also compare these probabilities by population principal strata defined by joint potential infection outcomes. In the next section, we define the probabilities of causation in a general setting. We then describe post-infection outcomes under the principal stratification framework. And finally, we present expressions for the probabilities of causation of post-infection outcomes in terms of principal strata-related parameters.

*Probabilities of causation*

To motivate our discussion, we consider a randomized vaccine trial with perfect compliance. Here, the outcome, $Y$, is infection-related hospitalization (1 = infection-related hospitalization, 0 = no infection-related hospitalization); under this definition, which is different from the one introduced in the next section, individuals without infection have $Y = 0$. For our analysis on causal attribution, the exposure of interest, $A$, corresponds to non-vaccination (1 = not vaccinated; 0 = vaccinated); based on this coding, an individual is exposed if she or he does not receive vaccination. Using this notation, the probability of necessity [4, 5] can be defined as:

$$p_N = \Pr(Y^0 = 0 | Y = 1, A = 1)$$

where $Y^a$ is the potential outcome under exposure level $a$. In our example, $p_N$ can be interpreted as the proportion of unvaccinated individuals admitted to hospital with infection who would not have had the outcome had they been vaccinated. For related work, see also the studies by Robins and Greenland [7], Yamamoto [3], and Dawid and colleagues [8].

Following Pearl [4, 5], we also consider the probabilities of sufficiency ($p_S$) and of necessity and sufficiency ($p_{NS}$):

$$p_S = \Pr(Y^1 = 1 | Y = 0, A = 0)$$

$$p_{NS} = \Pr(Y^1 = 1, Y^0 = 0)$$

Here, $p_S$ can be viewed as the proportion of vaccinated non-hospitalised individuals who would have been admitted to hospital with infection had they not been vaccinated. The probability of necessity and sufficiency, on the other hand, does not involve conditioning on observed events, and corresponds to the proportion of individuals for whom exposure would be necessary and sufficient.

*Post-infection outcomes and the principal stratification framework*

In analyses of post-infection outcomes, researchers explicitly account for infection status, and outcomes such as disease severity are only defined for individuals who acquire infection (here, denoted by the variable $I$). In our example, the corresponding post-infection outcome would be infection-related hospitalisation ($Y = 1$) versus infection that does not require hospitalisation ($Y = 0$); here, both levels of the variable $Y$ require presence of infection, and for participants with $I = 0$, $Y$ is undefined.

In such studies, since participants who acquire infection under different exposure levels might not represent a common set of individuals, the approach that compares outcomes by exposure status among those observed to be infected does not have a causal interpretation. However, Hudgens and Halloran [9] showed that the principal stratification framework can be used, in this context, to define a causal estimand. Specifically, investigators can estimate the effect of the exposure on the post-infection outcome for participants who would become infected regardless of exposure level. The application of this framework involves categorizing the population into principal strata defined by the joint potential infection outcomes, $I^1$ and $I^0$: $\{j|\ I_j^1 = 1, I_j^0 = 1\}$, $\{j|\ I_j^1 = 1, I_j^0 = 0\}$, $\{j|\ I_j^1 = 0, I_j^0 = 1\}$, $\{j|\ I_j^1 = 0, I_j^0 = 0\}$; $j$ denotes different study participants. The proportions of these strata are represented, respectively, by the parameters $\theta_{11}, \theta_{10}, \theta_{01}, \theta_{00}$. Further, for the "always infected" stratum (defined by $I^1 = 1$, $I^0 = 1$), the set of parameters $\{\emptyset_{11}, \emptyset_{10}, \emptyset_{01}, \emptyset_{00}\}$ corresponds to the proportions of the strata defined by the joint potential post-infection outcomes, with $\emptyset_{mn} = \Pr(Y^1 = m, Y^0 = n\ |I^1 = 1, I^0 = 1)$. Finally, for individuals for whom the exposure (here, non-vaccination) causes infection ($I^1 = 1, I^0 = 0$), $\gamma$ denotes the probability of the outcome under exposure, and for individuals with joint potential outcomes ($I^1 = 0, I^0 = 1$), $\beta$ denotes the probability of the outcome under absence of exposure; that is, $\gamma = \Pr(Y^1 = 1\ |I^1 = 1, I^0 = 0)$ and $\beta = \Pr(Y^0 = 1\ |I^1 = 0, I^0 = 1)$. **Table 1** shows the different response types defined based on potential outcomes of $I$ and $Y$.

**Table 1.** Potential infection and post-infection outcomes. Here, * indicates that $Y^a$ is not defined. This table is similar to that in the paper by Hudgens and Halloran [9]; however, the exposure here is absence of vaccination, rather than vaccination. See also the table in [10].

| | **Potential infection outcomes** | | | **Potential post-infection outcomes** | | |
|---|---|---|---|---|---|---|
| | $I^1$ | $I^0$ | *Proportions* | $Y^1$ | $Y^0$ | *Proportions* |
| *Always infected* | 1 | 1 | $\theta_{11}$ | 1 | 1 | $\theta_{11}\phi_{11}$ |
| *Always infected* | 1 | 1 | | 1 | 0 | $\theta_{11}\phi_{10}$ |
| *Always infected* | 1 | 1 | | 0 | 1 | $\theta_{11}\phi_{01}$ |
| *Always infected* | 1 | 1 | | 0 | 0 | $\theta_{11}\phi_{00}$ |
| *Causative* | 1 | 0 | $\theta_{10}$ | 0 | * | $\theta_{10}(1-\gamma)$ |
| *Causative* | 1 | 0 | | 1 | * | $\theta_{10}\gamma$ |
| *Defiers* | 0 | 1 | $\theta_{01}$ | * | 0 | $\theta_{01}(1-\beta)$ |
| *Defiers* | 0 | 1 | | * | 1 | $\theta_{01}\beta$ |
| *Never infected* | 0 | 0 | $\theta_{00}$ | * | * | $\theta_{00}$ |

*Probabilities of causation in the context of post-infection outcomes*

For observed post-infection outcomes, the probability of necessity should reflect the fact that individuals who would not develop the outcome in the counterfactual situation where exposure was absent would do so either because they would not become infected or because they would not require hospitalisation after infection. Analogously, causal sufficiency can be defined not only for individuals who were infected and did not require hospitalization (individuals with mild disease, $Y = 0$), but also for those who were not infected ($Y = *$). Thus, in this context, the probabilities of causation may be defined as:

$$p_N = \Pr(Y^0 = 0 \ or \ * \ | Y = 1, A = 1)$$

$$p_S = \Pr(Y^1 = 1 | Y = 0 \ or \ *, A = 0)$$

$$p_{NS} = \Pr(Y^1 = 1, Y^0 = 0 \ or \ *)$$

where $Y^a = *$ indicates that the potential post-infection outcome is undefined when $I^a = 0$ [6, 9], and $Y^a = 0 \ or \ *$ indicates that the potential post-infection outcome is either mild disease ($Y^a = 0$) or undefined ($Y^a = *$). Thus, in this type of study, both $Y^0 = 0$ and $Y^0 = *$ suggest causal necessity, and causal sufficiency is relevant for those with observed $Y = 0$ or $Y = *$.

Given these definitions, we now present expressions for these probabilities that use principal stratification parameters. The derivations are presented in the *Supplementary Appendix*; we assumed consistency (i.e., $I^a = I$ if $A = a$, and $Y^a = Y$ if $A = a$), the stable unit treatment value assumption and exchangeability, $(I^1, I^0, Y^1, Y^0) \perp\!\!\!\perp A$. We note, however, that although the stable unit treatment value assumption is often made in vaccine studies, as with other infectious disease studies, it might only hold approximately or not hold.

Firstly, the probability of necessity can be expressed as a weighted average of the corresponding probabilities in the "always infected" ($p_{N,AI}$) and the "causative" ($p_{N,C}$) strata:

$$p_N = p_{N,AI} \left( \frac{\theta_{11}(\emptyset_{11} + \emptyset_{10})}{(\theta_{11}(\emptyset_{11} + \emptyset_{10}) + \theta_{10}\gamma)} \right) + p_{N,C} \left( \frac{\theta_{10}\gamma}{(\theta_{11}(\emptyset_{11} + \emptyset_{10}) + \theta_{10}\gamma)} \right)$$

$$p_{N,AI} = \frac{\emptyset_{10}}{(\emptyset_{11} + \emptyset_{10})}$$

$$p_{N,C} = 1$$

where, for $p_N$, stratum-specific probabilities are weighted based on relative frequencies of the the two strata among exposed individuals with the outcome. Further, it follows from the definitions of the infection-based principal strata and the post-infection outcome that for individuals for whom exposure has an effect on infection ($I^1 = 1, I^0 = 0$), the probability of necessity, $p_{N,C}$, is 1. For the "always infected" stratum, $p_{N,AI}$ depends on the relative frequencies of individuals with the following potential outcomes, ($Y^1 = 1, Y^0 = 1, I^1 = 1, I^0 = 1$) and ($Y^1 = 1, Y^0 = 0, I^1 = 1, I^0 = 1$). Finally, notice that the first term in the expression for $p_N$, $p_{N,AI} \left( \frac{\theta_{11}(\emptyset_{11}+\emptyset_{10})}{(\theta_{11}(\emptyset_{11}+\emptyset_{10}) + \theta_{10}\gamma)} \right)$, quantifies, for individuals with unknown principal strata membership, the contribution of disease severity to causal necessity, and the second term, the contribution of infection acquisition.

The probability of sufficiency, in its turn, can be expressed as:

$$p_S = p_{S,AI} \left( \frac{\theta_{11}(\emptyset_{10} + \emptyset_{00})}{\theta_{11}(\emptyset_{10} + \emptyset_{00}) + \theta_{10} + \theta_{01}(1-\beta) + \theta_{00}} \right)$$
$$+ p_{S,C} \left( \frac{\theta_{10}}{\theta_{11}(\emptyset_{10} + \emptyset_{00}) + \theta_{10} + \theta_{01}(1-\beta) + \theta_{00}} \right)$$

$$p_{S,AI} = \frac{\emptyset_{10}}{(\emptyset_{10} + \emptyset_{00})}$$

$$p_{S,C} = \gamma$$

$$p_{S,D} = p_{S,NI} = 0$$

where $p_{S,AI}$ and $p_{S,C}$ represent the probabilities of sufficiency in the "always infected" and "causative" strata. As infection is a necessary step for $Y^1 = 1$, the probability of sufficiency for the "never infected" stratum, $p_{S,NI}$, and for individuals with the potential outcomes ($I^1 = 0, I^0 = 1$), $p_{S,D}$, is 0. In the expression for $p_S$, the denominator in the stratum-specific weights corresponds to the proportion of the population that would not develop the outcome if unexposed. Note also that, different from $p_N$, whether individuals were uninfected ($Y = *$) or had mild infection ($Y = 0$) is assumed to be observed, rather than a counterfactual outcome; thus, two additional probabilities can be defined, $p_{S-Mild} = \Pr(Y^1 = 1 | Y = 0, A = 0)$ and $p_{S-Uninf} = \Pr(Y^1 = 1 | Y = *, A = 0)$ (see *Supplementary Appendix* for further discussion).

Finally, the probabilities of necessity and sufficiency are:

$p_{NS} = p_{NS,AI}\, \theta_{11} + p_{NS,C}\, \theta_{10}$

$p_{NS,AI} = \emptyset_{10}$

$p_{NS,C} = \gamma$

$p_{NS,D} = p_{NS,NI} = 0$

In the *Supplementary Appendix*, we show that these expressions satisfy the Lemma 1 in [4], that describes the relation between the three probabilities:

$p_{NS} = p_N \Pr(Y = 1, A = 1) + p_S \Pr(Y = 0, A = 0)$

Moreover, although our focus is on post-infection outcomes, in the *Supplementary Appendix*, we also present expressions for the probabilities of causation of the infection outcome.

**Table 2** shows a numerical example with several scenarios. The exposure (non-vaccination) is assumed to increase the risk of both infection $((\theta_{11} + \theta_{10}) - (\theta_{11} + \theta_{01}) > 0)$, and the post-infection outcome in the "always infected" stratum (the principal effect $(\emptyset_{11} + \emptyset_{10}) - (\emptyset_{11} + \emptyset_{01}) > 0)$. As can be seen, for both the "always infected" stratum and the total population, all three probabilities would increase with a stronger principal effect, that is, with a stronger effect of the exposure on disease severity among those individuals who would develop infection regardless of exposure. Further, an increase in the magnitude of the effect on infection would also affect the probabilities in the total population but would not affect the probabilities in the principal strata.

Table 2. Probabilities of causation for scenarios where principal stratification parameters vary. For each scenario, parameters that varied relative to Scenario 1 are highlighted in grey. In Scenario 3, the principal effect in the "always infected" stratum was assumed to be stronger than the corresponding effect in Scenario 1; and in Scenario 4, the effect of the exposure on infection acquisition was assumed to be stronger than that of Scenario 1. For all scenarios, we assumed: $\theta_{00} = 0.15$, $\emptyset_{11} = 0.40$, $\emptyset_{00} = 0.10$, $\beta = 0.40$. In the upper part of the table, we show the data that would be observed in a clinical trial, and the bottom part of the table shows probabilities of causation. The quantities $\Pr(Y = 1|I = 1, A = 1)$ and $\Pr(Y = 1|I = 1, A = 0)$ can be calculated using the proportions of the principal strata for participants with $I = 1$ and the stratum-specific potential post-infection outcome parameters (see, for example, equations 15 and 17 in the paper by Hudgens and Halloran, although, there, the authors considered a protective exposure). Notice also that the crude comparison among those infected (two rightmost columns in the upper part of the table) would suggest a conclusion different from that based on the principal effect in the "always infected" stratum.

| | Parameters | | | | | | Observed data | | | |
|---|---|---|---|---|---|---|---|---|---|---|
| Scenarios | $\theta_{11}$ | $\theta_{10}$ | $\theta_{01}$ | $\emptyset_{10}$ | $\emptyset_{01}$ | $\gamma$ | $\Pr(I=1\|A=1)$ | $\Pr(I=1\|A=0)$ | $\Pr(Y=1\|A=1,I=1)$ | $\Pr(Y=1\|A=0,I=1)$ |
| 1 | 0.30 | 0.30 | 0.25 | 0.30 | 0.20 | 0.20 | 0.60 | 0.55 | 0.45 | 0.51 |
| 2 | 0.30 | 0.30 | 0.25 | 0.30 | 0.20 | 0.05 | 0.60 | 0.55 | 0.38 | 0.51 |
| 3 | 0.30 | 0.30 | 0.25 | 0.45 | 0.05 | 0.20 | 0.60 | 0.55 | 0.53 | 0.43 |
| 4 | 0.30 | 0.50 | 0.05 | 0.30 | 0.20 | 0.20 | 0.80 | 0.35 | 0.39 | 0.57 |

| | Parameters | | | | | | Probabilities of causation | | | | | | | | |
|---|---|---|---|---|---|---|---|---|---|---|---|---|---|---|---|
| Scenarios | $\theta_{11}$ | $\theta_{10}$ | $\theta_{01}$ | $\emptyset_{10}$ | $\emptyset_{01}$ | $\gamma$ | $p_N$ | $p_{N,AI}$ | $p_{N,C}$ | $p_S$ | $p_{S,AI}$ | $p_{S,C}$ | $p_{NS}$ | $p_{NS,AI}$ | $p_{NS,C}$ |
| 1 | 0.30 | 0.30 | 0.25 | 0.30 | 0.20 | 0.20 | 0.56 | 0.43 | 1 | 0.21 | 0.75 | 0.20 | 0.15 | 0.30 | 0.20 |
| 2 | 0.30 | 0.30 | 0.25 | 0.30 | 0.20 | 0.05 | 0.47 | 0.43 | 1 | 0.15 | 0.75 | 0.05 | 0.11 | 0.30 | 0.05 |
| 3 | 0.30 | 0.30 | 0.25 | 0.45 | 0.05 | 0.20 | 0.62 | 0.53 | 1 | 0.25 | 0.82 | 0.20 | 0.20 | 0.45 | 0.20 |
| 4 | 0.30 | 0.50 | 0.05 | 0.30 | 0.20 | 0.20 | 0.61 | 0.43 | 1 | 0.24 | 0.75 | 0.20 | 0.19 | 0.30 | 0.20 |

*Conclusion*

Here, we considered a type of causal analysis that does not aim to estimate effects but rather aims to quantify causal attribution. Our derivations, and the resulting expressions, show that in the context of post-infection outcomes, the probabilities of causation can be ascribed to different mechanisms: for instance, $p_N$ depends on the probability of the counterfactual outcome that leads to mild infection, and the probability of the counterfactual outcome that leads to absence of infection. In settings where causal attribution is important, investigators might want to discriminate between, and consider specific implications of, these processes. For example, as illustrated in **Table 2**, a stronger effect on disease severity in the stratum of individuals who would become infected regardless of exposure would imply higher probabilities of causation in the entire population.

In general, however, neither the probabilities of causation nor the principal stratification parameters are identifiable without strong assumptions. Indeed, Hudgens and Halloran showed (see Section 3 in [9]) that the stable unit treatment value assumption, exchangeability, and mononicity are not sufficient to identify the principal effect of interest, and proposed selection models for the estimation of these effects. As to the probabilities, Pearl [4], and Tian and Pearl [11] described bounds based on experimental data and combinations of experimental and observational data (see also [12-14]); in fact, if one does not account for infection acquisition when defining the outcome, these bounds could be directly applied. Furthermore, Kuroki and Cai [15] described analyses of probabilities of causation using covariate information, and derived bounds for these probabilities using covariate data (see also [16]). Note that although we focused on binary outcomes, Li and Pearl [17] provided bounds for probabilities of causation in settings where treatment and outcome are not binary (see Theorem 7 of the paper by Li and Pearl; see also [18]); such an approach could be used, for example, when different degrees of disease severity are considered (e.g., hospital admission not requiring intensive care and hospital admission that required intensive care). Interest may also lie in stratum-specific probabilities, and in this case, further work is needed so that informative bounds can be estimated for these probabilities (relatedly, see work by Yamamoto [3], that describes identification of the probability of necessity [the term used there is probability of causal attribution] in a principal stratum using instrumental variables).

**Funding**

This research did not receive any specific grant from funding agencies in the public, commercial, or not-for-profit sectors.

**Conflict of Interest**

The author has no conflict of interest to declare.

**Data availability statement**

Not applicable.

# References

1. Rubin, D.B., *Causal Inference Using Potential Outcomes: Design, Modeling, Decisions.* Journal of the American Statistical Association, 2005. **100**(469): p. 322–331.
2. Hernan, M.A. and J.M. Robins, *Estimating causal effects from epidemiological data.* J Epidemiol Community Health, 2006. **60**(7): p. 578-86.
3. Yamamoto, T., *Understanding the Past: Statistical Analysis of Causal Attribution.* American Journal of Political Science, 2011. **56**(1): p. 237-256.
4. Pearl, J., *Probabilities Of Causation: Three Counterfactual Interpretations And Their Identification.* Synthese, 1999. **121**: p. 93-149.
5. Pearl, J., *Causality: Models, Reasoning, and Inference*. 2nd Edition ed. 2009: Cambridge University Press.
6. Zhang, J.L. and D.B. Rubin, *Estimation of Causal Effects via Principal Stratification When Some Outcomes Are Truncated by "Death."*. Journal of Educational and Behavioral Statistics, 2003. **28**(4): p. 353–368.
7. Robins, J. and S. Greenland, *The probability of causation under a stochastic model for individual risk.* Biometrics, 1989. **45**(4): p. 1125-38.
8. Dawid, A.P., M. Musio, and R. Murtas, *The probability of causation.* Law, Probability and Risk, 2017. **16**(4): p. 163–179.
9. Hudgens, M.G. and M.E. Halloran, *Causal Vaccine Effects on Binary Postinfection Outcomes.* J Am Stat Assoc, 2006. **101**(473): p. 51-64.
10. Goncalves, B.P. and E. Suzuki, *Preventable Fraction in the Context of Disease Progression.* Epidemiology, 2024. **35**(6): p. 801-804.
11. Tian, J. and J. Pearl, *Probabilities of causation: Bounds and identification.* Annals of Mathematics and Artificial Intelligence, 2000. **28**: p. 287–313.
12. Mueller, S., A. Li, and J. Pearl, *Causes of Effects: Learning Individual Responses from Population Data.* Proceedings of the Thirty-First International Joint Conference on Artificial Intelligence, 2022: p. 2712-2718.
13. Li, A. and J. Pearl, *Probabilities of Causation: Role of Observational Data.* Proceedings of the 26th International Conference on Artificial Intelligence and Statistics, 2023.
14. Peña, J.M., *Bounding the probabilities of benefit and harm through sensitivity parameters and proxies.* Journal of Causal Inference, 2023. **11**: p. 20230012.
15. Kuroki, M. and Z. Cai, *Statistical Analysis of 'Probabilities of Causation' Using Co-Variate Information.* Scandinavian Journal of Statistics, 2011. **38**(3): p. 564-577.

16. Shingaki, R. and M. Kuroki. *Identification and Estimation of "Causes of Effects" using Covariate-Mediator Information*. in *Proceedings of The 27th International Conference on Artificial Intelligence and Statistics*. 2024. Proceedings of Machine Learning Research.
17. Li, A. and J. Pearl. *Probabilities of Causation with Nonbinary Treatment and Effect*. in *Proceedings of the AAAI Conference on Artificial Intelligence*. 2024.
18. Kawakami, Y., M. Kuroki, and J. Tian. *Probabilities of Causation for Continuous and Vector Variables*. in *Proceedings of the Fortieth Conference on Uncertainty in Artificial Intelligence*. 2024. Proceedings of Machine Learning Research.

*Supplementary Appendix*

**Title**

Probabilities of causation and post-infection outcomes

**Authors**

Bronner P. Gonçalves[1]

**Affiliations**

[1] Faculty of Health and Medical Sciences, University of Surrey, 30 Priestley Rd, Guildford GU2 7YH, United Kingdom

**Table of contents**

*Probabilities of causation and the infection outcome*

Here, under consistency, stable unit treatment value assumption (SUTVA) and exchangeability, $(I^1, I^0) \perp\!\!\!\perp A$, we derive the probabilities of causation for the infection outcome. The principal strata-related parameters $\theta_{11}$, $\theta_{10}$ and $\theta_{00}$ were defined in the main text. Note that, different from the corresponding expression for the post-infection outcome, the expression for the probability of sufficiency for the infection outcome, $p_{S-I}$, does not include the parameter $\theta_{01}$, that corresponds to the proportion of the population in the "defiers" principal stratum (that is, the stratum with joint potential outcomes $I^1 = 0, I^0 = 1$).

*Probability of necessity*

$$p_{N-I} = \Pr(I^0 = 0 \mid I = 1, A = 1) = \frac{\theta_{10}}{\theta_{11} + \theta_{10}}$$

Proof

$$p_{N-I} = \Pr(I^0 = 0 \mid I = 1, A = 1) = \frac{\Pr(I^0 = 0, I = 1, A = 1)}{\Pr(I = 1, A = 1)}$$

$$= \frac{\Pr(I^0 = 0, I^1 = 1, A = 1)}{\Pr(I^1 = 1, A = 1)} \quad (consistency)$$

$$= \frac{\Pr(I^0 = 0, I^1 = 1 \mid A = 1) \Pr(A = 1)}{\Pr(I^1 = 1 \mid A = 1) \Pr(A = 1)}$$

$$= \frac{\Pr(I^0 = 0, I^1 = 1)}{\Pr(I^1 = 1)} \quad (exchangeability)$$

$$= \frac{\theta_{10}}{\theta_{11} + \theta_{10}}$$

*Probability of sufficiency*

$$p_{S-I} = \Pr(I^1 = 1 \mid I = 0, A = 0) = \frac{\theta_{10}}{\theta_{10} + \theta_{00}}$$

Proof

$$p_{S-I} = \Pr(I^1 = 1 \mid I = 0, A = 0) = \frac{\Pr(I^1 = 1, I = 0, A = 0)}{\Pr(I = 0, A = 0)}$$

$$= \frac{\Pr(I^1 = 1, I^0 = 0, A = 0)}{\Pr(I^0 = 0, A = 0)} \quad (consistency)$$

$$= \frac{\Pr(I^1 = 1, I^0 = 0 \mid A = 0) \Pr(A = 0)}{\Pr(I^0 = 0 \mid A = 0) \Pr(A = 0)}$$

$$= \frac{\Pr(I^1 = 1, I^0 = 0)}{\Pr(I^0 = 0)} \quad (exchangeability)$$

$$= \frac{\theta_{10}}{\theta_{10} + \theta_{00}}$$

*Probability of necessity and sufficiency*

$$p_{NS-I} = \Pr(I^1 = 1, I^0 = 0) = \theta_{10}$$

This follows from the definition of the parameter $\theta_{10}$.

*Probabilities of causation and the post-infection outcome*

In this section, we derive, under consistency, SUTVA and exchangeability, $(I^1, I^0, Y^1, Y^0) \perp\!\!\!\perp A$, expressions for the probabilities of causation of the post-infection outcome.

<u>Probability of necessity and sufficiency</u>

$p_{NS} = \Pr(Y^1 = 1, Y^0 = 0 \text{ or } *)$

$= \Pr(Y^1 = 1, Y^0 = 0 \text{ or } * | I^1 = 1, I^0 = 1) \Pr(I^1 = 1, I^0 = 1) + \Pr(Y^1 = 1, Y^0 = 0 \text{ or } * | I^1 = 1, I^0 = 0) \Pr(I^1 = 1, I^0 = 0)$
$\quad + \Pr(Y^1 = 1, Y^0 = 0 \text{ or } * | I^1 = 0, I^0 = 1) \Pr(I^1 = 0, I^0 = 1)$
$\quad + \Pr(Y^1 = 1, Y^0 = 0 \text{ or } * | I^1 = 0, I^0 = 0) \Pr(I^1 = 0, I^0 = 0) \quad \text{(law of total probability)}$

$= \Pr(Y^1 = 1, Y^0 = 0 \text{ or } * | I^1 = 1, I^0 = 1) \Pr(I^1 = 1, I^0 = 1) + \Pr(Y^1 = 1, Y^0 = 0 \text{ or } * | I^1 = 1, I^0 = 0) \Pr(I^1 = 1, I^0 = 0)$
$\quad + 0 \times \Pr(I^1 = 0, I^0 = 1) + 0 \times \Pr(I^1 = 0, I^0 = 0)$

$= \Pr(Y^1 = 1, Y^0 = 0 \text{ or } * | I^1 = 1, I^0 = 1) \, \theta_{11} + \Pr(Y^1 = 1, Y^0 = 0 \text{ or } * | I^1 = 1, I^0 = 0) \, \theta_{10}$

$= \emptyset_{10} \theta_{11} + \gamma \theta_{10}$

*Probability of necessity and sufficiency in the different principal strata*

$p_{NS,AI} = \emptyset_{10}$

$p_{NS,C} = \gamma$

$p_{NS,D} = p_{NS,NI} = 0$

where $p_{NS,AI}, p_{NS,C}, p_{NS,D}, p_{NS,NI}$ correspond to the probabilities of necessity and sufficiency in the "always infected", "causative", "defiers" and "never infected" strata (respectively, $\{j|\ I_j^1 = 1, I_j^0 = 1\}$, $\{j|\ I_j^1 = 1, I_j^0 = 0\}$, $\{j|\ I_j^1 = 0, I_j^0 = 1\}$, $\{j|\ I_j^1 = 0, I_j^0 = 0\}$, where $j$ denotes study participants).

*Probability of necessity*

$p_N = \Pr(Y^0 = 0 \text{ or } * | Y = 1, A = 1)$

$= \Pr(Y^0 = 0 \text{ or } * | Y = 1, A = 1, (I^1 = 1, I^0 = 1)) \Pr(I^1 = 1, I^0 = 1 | Y = 1, A = 1) + \Pr(Y^0 = 0 \text{ or } * | Y = 1, A = 1, (I^1 = 1, I^0 = 0)) \Pr(I^1 = 1, I^0 = 0 | Y = 1, A = 1) + \Pr(Y^0 = 0 \text{ or } * | Y = 1, A = 1, (I^1 = 0, I^0 = 1)) \Pr(I^1 = 0, I^0 = 1 | Y = 1, A = 1) + \Pr(Y^0 = 0 \text{ or } * | Y = 1, A = 1, (I^1 = 0, I^0 = 0)) \Pr(I^1 = 0, I^0 = 0 | Y = 1, A = 1)$    *(law of total probability for conditional probabilities)*

$= \Pr(Y^0 = 0 \text{ or } * | Y = 1, A = 1, (I^1 = 1, I^0 = 1)) \Pr(I^1 = 1, I^0 = 1 | Y = 1, A = 1) + \Pr(Y^0 = 0 \text{ or } * | Y = 1, A = 1, (I^1 = 1, I^0 = 0)) \Pr(I^1 = 1, I^0 = 0 | Y = 1, A = 1) + 0 + 0$    *(consistency and outcome definition)*

$= \Pr(Y^0 = 0 \text{ or } * | Y = 1, A = 1, (I^1 = 1, I^0 = 1)) \left( \frac{\theta_{11}(\emptyset_{11}+\emptyset_{10})}{(\theta_{11}(\emptyset_{11}+\emptyset_{10}) + \theta_{10}\gamma)} \right) + \Pr(Y^0 = 0 \text{ or } * | Y = 1, A = 1, (I^1 = 1, I^0 = 0)) \left( \frac{\theta_{10}\gamma}{(\theta_{11}(\emptyset_{11}+\emptyset_{10}) + \theta_{10}\gamma)} \right)$

$= \Pr(Y^0 = 0 \text{ or } * | Y^1 = 1, (I^1 = 1, I^0 = 1)) \left( \frac{\theta_{11}(\emptyset_{11}+\emptyset_{10})}{(\theta_{11}(\emptyset_{11}+\emptyset_{10}) + \theta_{10}\gamma)} \right) + \Pr(Y^0 = 0 \text{ or } * | Y^1 = 1, (I^1 = 1, I^0 = 0)) \left( \frac{\theta_{10}\gamma}{(\theta_{11}(\emptyset_{11}+\emptyset_{10}) + \theta_{10}\gamma)} \right)$

*(consistency and exchangeability [and the decomposition and weak union properties])*

$$= \frac{\emptyset_{10}}{(\emptyset_{11} + \emptyset_{10})} \left( \frac{\theta_{11}(\emptyset_{11} + \emptyset_{10})}{(\theta_{11}(\emptyset_{11} + \emptyset_{10}) + \theta_{10}\gamma)} \right) + 1 \text{ x} \left( \frac{\theta_{10}\gamma}{(\theta_{11}(\emptyset_{11} + \emptyset_{10}) + \theta_{10}\gamma)} \right)$$

$$= \frac{\theta_{11}\emptyset_{10} + \theta_{10}\gamma}{(\theta_{11}(\emptyset_{11} + \emptyset_{10}) + \theta_{10}\gamma)}$$

Notice that above, under consistency and exchangeability, we used the fact that

$$\Pr(Y = 1, A = 1) = \Pr(Y^1 = 1, A = 1)$$

$$= \Pr(Y^1 = 1 | A = 1) \Pr(A = 1)$$

$$= \Pr(Y^1 = 1) \Pr(A = 1)$$

$$= (\theta_{11}(\emptyset_{11} + \emptyset_{10}) + \theta_{10}\gamma) \Pr(A = 1)$$

The expression for the term $\Pr(I^1 = 1, I^0 = 1 | Y = 1, A = 1)$ was then obtained:

$$\Pr(I^1 = 1, I^0 = 1 | Y = 1, A = 1) = \frac{\Pr(I^1 = 1, I^0 = 1, Y = 1, A = 1)}{\Pr(Y = 1, A = 1)}$$

$$= \frac{\Pr(I^1 = 1, I^0 = 1, Y^1 = 1, A = 1)}{\Pr(Y = 1, A = 1)} \quad (consistency)$$

$$= \frac{\Pr(I^1 = 1, I^0 = 1, Y^1 = 1)\Pr(A = 1)}{\Pr(Y = 1, A = 1)} \quad (exchangeability)$$

$$= \frac{\theta_{11}(\emptyset_{11} + \emptyset_{10})}{(\theta_{11}(\emptyset_{11} + \emptyset_{10}) + \theta_{10}\gamma)}$$

The expression for $\Pr(I^1 = 1, I^0 = 0 | Y = 1, A = 1)$ was obtained in a similar manner.

*Probability of necessity in the different principal strata*

The expressions for the "always infected" and "causative" strata are:

$$p_{N,AI} = \frac{\emptyset_{10}}{(\emptyset_{11} + \emptyset_{10})}$$

$$p_{N,C} = 1$$

As $\Pr(Y = 1, A = 1, (I^1 = 0, I^0 = 1)) = 0$ and $\Pr(Y = 1, A = 1, (I^1 = 0, I^0 = 0)) = 0$, the probability of necessity is not defined for the "defiers" and "never infected" strata.

*Probability of sufficiency*

$p_S = \Pr(Y^1 = 1 | Y = 0 \text{ or } *, A = 0)$

$= \Pr(Y^1 = 1 | Y = 0 \text{ or } *, A = 0, (I^1 = 1, I^0 = 1)) \Pr(I^1 = 1, I^0 = 1 | Y = 0 \text{ or } *, A = 0) + \Pr(Y^1 = 1 | Y = 0 \text{ or } *, A = 0, (I^1 = 1, I^0 = 0)) \Pr(I^1 = 1, I^0 = 0 | Y = 0 \text{ or } *, A = 0) + \Pr(Y^1 = 1 | Y = 0 \text{ or } *, A = 0, (I^1 = 0, I^0 = 1)) \Pr(I^1 = 0, I^0 = 1 | Y = 0 \text{ or } *, A = 0) + \Pr(Y^1 = 1 | Y = 0 \text{ or } *, A = 0, (I^1 = 0, I^0 = 0)) \Pr(I^1 = 0, I^0 = 0 | Y = 0 \text{ or } *, A = 0)$     (*law of total probability for conditional probabilities*)

$= \Pr(Y^1 = 1 | Y = 0 \text{ or } *, A = 0, (I^1 = 1, I^0 = 1)) \Pr(I^1 = 1, I^0 = 1 | Y = 0 \text{ or } *, A = 0) + \Pr(Y^1 = 1 | Y = 0 \text{ or } *, A = 0, (I^1 = 1, I^0 = 0)) \Pr(I^1 = 1, I^0 = 0 | Y = 0 \text{ or } *, A = 0) + 0 \text{ x } \Pr(I^1 = 0, I^0 = 1 | Y = 0 \text{ or } *, A = 0) + 0 \text{ x } \Pr(I^1 = 0, I^0 = 0 | Y = 0 \text{ or } *, A = 0)$

$= \Pr(Y^1 = 1 | Y^0 = 0 \text{ or } *, (I^1 = 1, I^0 = 1)) \Pr(I^1 = 1, I^0 = 1 | Y = 0 \text{ or } *, A = 0) + \Pr(Y^1 = 1 | Y^0 = 0 \text{ or } *, (I^1 = 1, I^0 = 0)) \Pr(I^1 = 1, I^0 = 0 | Y = 0 \text{ or } *, A = 0)$

(*consistency and exchangeability [and the decomposition and weak union properties]*)

$$= \Pr(Y^1 = 1 | Y^0 = 0 \text{ or } *, (I^1 = 1, I^0 = 1)) \left( \frac{\theta_{11}(\emptyset_{10} + \emptyset_{00})}{\theta_{11}(\emptyset_{10} + \emptyset_{00}) + \theta_{10} + \theta_{01}(1 - \beta) + \theta_{00}} \right)$$

$$+ \Pr(Y^1 = 1 | Y^0 = 0 \text{ or } *, (I^1 = 1, I^0 = 0)) \left( \frac{\theta_{10}}{\theta_{11}(\emptyset_{10} + \emptyset_{00}) + \theta_{10} + \theta_{01}(1 - \beta) + \theta_{00}} \right)$$

$$= \frac{\emptyset_{10}}{(\emptyset_{10} + \emptyset_{00})} \frac{\theta_{11}(\emptyset_{10} + \emptyset_{00})}{\theta_{11}(\emptyset_{10} + \emptyset_{00}) + \theta_{10} + \theta_{01}(1 - \beta) + \theta_{00}} + \gamma \frac{\theta_{10}}{\theta_{11}(\emptyset_{10} + \emptyset_{00}) + \theta_{10} + \theta_{01}(1 - \beta) + \theta_{00}}$$

$$= \frac{\theta_{11} \emptyset_{10} + \theta_{10} \gamma}{\theta_{11}(\emptyset_{10} + \emptyset_{00}) + \theta_{10} + \theta_{01}(1 - \beta) + \theta_{00}}$$

Above, under consistency and exchangeability, we used the fact that

$$\Pr(Y = 0 \text{ or } *, A = 0) = \Pr(Y^0 = 0 \text{ or } *, A = 0)$$

$$= \Pr(Y^0 = 0 \text{ or } * | A = 0) \Pr(A = 0)$$

$$= \Pr(Y^0 = 0 \text{ or } *) \Pr(A = 0)$$

$$= (\theta_{11}(\emptyset_{10} + \emptyset_{00}) + \theta_{10} + \theta_{01}(1 - \beta) + \theta_{00}) \Pr(A = 0)$$

The expression for the term $\Pr(I^1 = 1, I^0 = 1 | Y = 0 \text{ or } *, A = 0)$ was then obtained:

$$\Pr(I^1 = 1, I^0 = 1 | Y = 0 \text{ or } *, A = 0) = \frac{\Pr(I^1 = 1, I^0 = 1, Y = 0 \text{ or } *, A = 0)}{\Pr(Y = 0 \text{ or } *, A = 0)}$$

$$= \frac{\Pr(I^1 = 1, I^0 = 1, Y^0 = 0 \text{ or } *, A = 0)}{\Pr(Y = 0 \text{ or } *, A = 0)} \quad (consistency)$$

$$= \frac{\Pr(I^1 = 1, I^0 = 1, Y^0 = 0 \text{ or } *) \Pr(A = 0)}{\Pr(Y = 0 \text{ or } *, A = 0)} \quad (exchangeability)$$

$$= \frac{\theta_{11}(\emptyset_{10} + \emptyset_{00})}{(\theta_{11}(\emptyset_{10} + \emptyset_{00}) + \theta_{10} + \theta_{01}(1 - \beta) + \theta_{00})}$$

$\Pr(I^1 = 1, I^0 = 0 | Y = 0 \text{ or } *, A = 0)$ was obtained in a similar manner.

*Probability of sufficiency in the different principal strata*

$$p_{S,AI} = \frac{\emptyset_{10}}{(\emptyset_{10} + \emptyset_{00})}$$

$$p_{S,C} = \gamma$$

$$p_{S,D} = p_{S,NI} = 0$$

Further, as mentioned in the main text, whether individuals were uninfected or had mild infection is observed; two additional probabilities can then be defined, $p_{S-Mild} = \Pr(Y^1 = 1|Y = 0, A = 0)$ and $p_{S-Uninf} = \Pr(Y^1 = 1|Y = *, A = 0)$. Below, we present expressions for these quantities in terms of principal stratification parameters:

$$p_{S-Mild} = \Pr(Y^1 = 1|Y = 0, A = 0)$$

$= \Pr(Y^1 = 1|Y = 0, A = 0, (I^1 = 1, I^0 = 1)) \Pr(I^1 = 1, I^0 = 1|Y = 0, A = 0) + \Pr(Y^1 = 1|Y = 0, A = 0, (I^1 = 0, I^0 = 1)) \Pr(I^1 = 0, I^0 = 1|Y = 0, A = 0)$ *(law of total probability for conditional probabilities)*

$= \Pr(Y^1 = 1|Y = 0, A = 0, (I^1 = 1, I^0 = 1)) \Pr(I^1 = 1, I^0 = 1|Y = 0, A = 0) + 0 \times \Pr(I^1 = 0, I^0 = 1|Y = 0, A = 0)$

$$= \frac{\emptyset_{10}}{(\emptyset_{10} + \emptyset_{00})} \frac{\theta_{11}(\emptyset_{10} + \emptyset_{00})}{\theta_{11}(\emptyset_{10} + \emptyset_{00}) + \theta_{01}(1 - \beta)}$$

$$= \frac{\theta_{11}\emptyset_{10}}{\theta_{11}(\emptyset_{10} + \emptyset_{00}) + \theta_{01}(1 - \beta)}$$

And,

$$p_{S-Uninf} = \Pr(Y^1 = 1 | Y = *, A = 0)$$

$$= \Pr(Y^1 = 1 | Y = *, A = 0, (I^1 = 1, I^0 = 0)) \Pr(I^1 = 1, I^0 = 0 | Y = *, A = 0) + \Pr(Y^1 = 1 | Y = *, A = 0, (I^1 = 0, I^0 = 0)) \Pr(I^1 = 0, I^0 = 0 | Y = *, A = 0) \quad (law\ of\ total\ probability\ for\ conditional\ probabilities)$$

$$= \Pr(Y^1 = 1 | Y = *, A = 0, (I^1 = 1, I^0 = 0)) \Pr(I^1 = 1, I^0 = 0 | Y = *, A = 0) + 0 \times \Pr(I^1 = 0, I^0 = 0 | Y = *, A = 0)$$

$$= \gamma \frac{\theta_{10}}{\theta_{10} + \theta_{00}}$$

$$= \frac{\theta_{10}\gamma}{\theta_{10} + \theta_{00}}$$

*Relation between probabilities*

The following relation holds between the probabilities of causation (Lemma 1 in [1]):

$$p_{NS} = p_N \Pr(Y = 1, A = 1) + p_S \Pr(Y = 0, A = 0)$$

In this section, we show that the relation holds for the expressions derived here; for the "defiers" and "never infected" strata, the probabilities of necessity and sufficiency and of sufficiency are 0.

*Total population*

$$p_{NS} = p_N \Pr(Y = 1, A = 1) + p_S \Pr(Y = 0 \text{ or } *, A = 0)$$

$$\theta_{11}\emptyset_{10} + \theta_{10}\gamma = \frac{\theta_{11}\emptyset_{10} + \theta_{10}\gamma}{(\theta_{11}(\emptyset_{11}+\emptyset_{10})+ \theta_{10}\gamma)} \Pr(A = 1)(\theta_{11}(\emptyset_{11} + \emptyset_{10}) + \theta_{10}\gamma) + \frac{\theta_{11}\emptyset_{10} + \theta_{10}\gamma}{(\theta_{11}(\emptyset_{10}+\emptyset_{00})+\theta_{10}+\theta_{01}(1-\beta)+\theta_{00})}(1 - \Pr(A = 1))(\theta_{11}(\emptyset_{10} + \emptyset_{00}) + \theta_{10} + \theta_{01}(1 - \beta) + \theta_{00})$$

$$\theta_{11}\emptyset_{10} + \theta_{10}\gamma = (\theta_{11}\emptyset_{10} + \theta_{10}\gamma)(\Pr(A = 1) + 1 - \Pr(A = 1))$$

$$\theta_{11}\emptyset_{10} + \theta_{10}\gamma = \theta_{11}\emptyset_{10} + \theta_{10}\gamma$$

For the second equality, consistency and exchangeability were used to show that

$$\Pr(Y = 1, A = 1) = \Pr(A = 1)(\theta_{11}(\emptyset_{11} + \emptyset_{10}) + \theta_{10}\gamma)$$

and

$$\Pr(Y = 0 \text{ or } *, A = 0) = (1 - \Pr(A = 1))(\theta_{11}(\emptyset_{10} + \emptyset_{00}) + \theta_{10} + \theta_{01}(1 - \beta) + \theta_{00})$$

For example,

$$\Pr(Y = 1, A = 1) = \Pr(Y^1 = 1, A = 1) \quad (consistency)$$

$$= \Pr(Y^1 = 1 | A = 1) \Pr(A = 1)$$

$$= \Pr(Y^1 = 1) \Pr(A = 1) \quad (exchangeability)$$

$$= (\theta_{11}(\emptyset_{11} + \emptyset_{10}) + \theta_{10}\gamma) \Pr(A = 1)$$

*"Always infected" stratum*

$$p_{NS,AI} = p_{N,AI} \Pr(Y = 1, A = 1|I^1 = 1, I^0 = 1)$$
$$+ p_{S,AI} \Pr(Y = 0 \text{ or } *, A = 0|I^1 = 1, I^0 = 1)$$

$$\emptyset_{10} = \frac{\emptyset_{10}}{(\emptyset_{11} + \emptyset_{10})} \Pr(A = 1)(\emptyset_{11} + \emptyset_{10}) + \frac{\emptyset_{10}}{(\emptyset_{10} + \emptyset_{00})}(1 - \Pr(A = 1))(\emptyset_{10} + \emptyset_{00})$$

$$\emptyset_{10} = \emptyset_{10}(\Pr(A = 1) + 1 - \Pr(A = 1))$$

$$\emptyset_{10} = \emptyset_{10}$$

Consistency and exchangeability were used to show

$$\Pr(Y = 1, A = 1|I^1 = 1, I^0 = 1) = \Pr(A = 1)(\emptyset_{11} + \emptyset_{10})$$

and

$$\Pr(Y = 0 \text{ or } *, A = 0 |I^1 = 1, I^0 = 1) = (1 - \Pr(A = 1))(\emptyset_{10} + \emptyset_{00})$$

*"Causative" stratum*

$$p_{NS,C} = p_{N,C} \Pr(Y = 1, A = 1|I^1 = 1, I^0 = 0) + p_{S,C} \Pr(Y = 0 \text{ or } *, A = 0|I^1 = 1, I^0 = 0)$$

$$\gamma = 1 \times \Pr(A = 1)\gamma + \gamma(1 - \Pr(A = 1))$$

$$\gamma = \gamma(\Pr(A = 1) + 1 - \Pr(A = 1))$$

$$\gamma = \gamma$$

# References

1. Pearl, J., *Probabilities Of Causation: Three Counterfactual Interpretations And Their Identification.* Synthese, 1999. **121**: p. 93-149.